\documentclass[preprint2]{aastex}

\slugcomment{manuscript}

\shorttitle{MODS V: Surface Brightness}
\shortauthors{Ichikawa et al.}

\begin{document}

\title{MOIRCS Deep Survey V: A Universal Relation for Stellar Mass and Surface 
Brightness of Galaxies}

\author{Takashi Ichikawa, Masaru Kajisawa, Toru Yamada, Masayuki Akiyama, Tomohiro Yoshikawa}
\affil{Astronomical Institute, Tohoku University, Aoba, Sendai 980-8578, Japan}
\email{ichikawa@astr.tohoku.ac.jp}
\author{Masato Onodera}
\affil{CEA-Saclay, DSM/DAPNIA/Service d'Astrophysique, 91191 Gif-sur-Yvette Cedex, France}
\and
\author{Masahiro Konishi}
\affil{Institute of Astronomy, University of Tokyo, Mitaka, Tokyo 181-0015, Japan}

\begin{abstract}
We present a universal linear correlation between the stellar mass and surface brightness (SB) of 
galaxies at $0.3<z<3$, using a deep $K$-band selected catalog in the GOODS-North region. 
The correlation has a nearly constant slope, independent of redshift and color of galaxies 
in the rest-$z$ frame.
Considering unresolved compact galaxies, the tight correlation 
gives a lower boundary of SB for a given stellar mass; lower SB galaxies are prohibited 
over the boundary.
The universal slope suggests that the stellar mass in galaxies was build up over 
their cosmic histories in a similar manner irrelevant to galaxy mass,
as oppose to the scenario that massive galaxies mainly accumulated their stellar mass by 
major merging.
In contrast, SB shows a strong dependence on redshift for a given stellar mass.
It evolves as $\sim(1+z)^{-2.0\sim-0.8}$, in addition to dimming as $(1+z)^4$ by 
the cosmological expansion effect.
The brightening depends on galaxy color and stellar mass.
The blue population (rest-frame $U-V<0$), which is dominated by young and star-forming galaxies, 
evolves  as $\sim(1+z)^{-0.8\pm0.3}$ in the rest-$V$ band.
On the other hand, the red population ($U-V>0$) and the massive galaxies ($M_*>10^{10}M_\sun$)
shows stronger brightening, $(1+z)^{-1.5\pm0.1}$.
Based on the comparison with galaxy evolution models, we find that the phenomena are 
well explained by the pure luminosity evolution of galaxies out to $z\sim3$.

\end{abstract}


\section{Introduction}

Galaxy evolution from high-$z$  to the present time is one of the central 
issues to be understood in $\Lambda$CDM paradigm.
Among many parameters to characterize galaxies, stellar mass, which is 
accumulated in a galaxy with aging, is one of 
the most fundamental quantities for understanding the star-forming and merging histories
of galaxies.
The surface brightness (SB) of galaxies is also key to studying 
galaxy structure. 
For example, Kormendy (1977) showed a tight correlation between SB and radius parameters 
for compact and normal galaxies.
Since then, many authors have studied the evolution of SB in the high-$z$ universe
(e.g., Jorgensen et al. 1995; Roche et al. 1998; Lilly et al. 1998; La Barbera et al. 2004; Barden et al. 2005; 
Ferreras et al. 2009; Trujillo 2004, 2006; Cimatti et al.\ 2008; van Dokkum et al.\ 2008; 
Muzzin et al.\ 2009; Damjanov et al.\ 2009).
The SB and its correlation with other structural parameters of local galaxies were 
investigated with SDSS galaxies (e.g., Shen et al.\ 2003; Bernardi et al.\ 2003).
A direct evidence of SB dimming as $(1+z)^{4}$ 
also intrigues us as a cosmological expansion test (Tolman 1930; Lubin \& Sandage 2001).

By analogy with the Kormendy relation, the correlation of stellar 
mass with SB of galaxies (hereafter called the stellar-mass Kormendy relation)
in the high-$z$ universe will give us a clue to understanding the history of 
stellar mass built up in galaxies over cosmic time.
In that context, we will study the correlation and its evolution
from $z\sim3$ to $z\sim0.3$, using a deep $K$-selected galaxy catalog.

In \S2, we describe the catalog we used. The data analysis and 
the result for the stellar-mass Kormendy relation and its evolution are 
detailed in \S3 and  \S4. The results are discussed in \S5.
We use the AB magnitude system, except for Vega $U-V$ color.

\section{Data}

We use the $K$-band selected catalog of the MOIRCS Deep Survey (MODS) in the GOODS--North region 
(Kajisawa et al. 2009, hereafter K09),
 which is based on our imaging observations in $JHKs$ bands with MOIRCS 
 (Suzuki et al. 2008) and archived data.
Four MOIRCS pointings cover 70\% of the GOODS--North region (103 arcmin$^2$, 
hereafter referred as `wide' field). 
One of the four pointings, which includes HDF-N (Williams et al. 1996), is  
the ultra-deep field of MODS (28 arcmin$^2$, `deep' field).
The deep imaging gives us SB data in a wide dynamic range,
so that it is also advantageous for studying low SB galaxies at high redshift, where
SB dimming due to the cosmological expansion is significant.

Although the detection completeness ($90\%$) of the catalog is $K\sim25$ mag for 
the wide field and $\sim 26$ mag for the deep field, we include fainter galaxies,
because the $K$-band limit does not concern the present study.
In fact, the sample  with $K=25$ limit gives the same result within error.
Instead, we will take account of the size and SB limits of galaxies.
The FWHMs of the final stacked images are $0\arcsec\!.46$ for the deep image
and $0\arcsec\!.53$--$0\arcsec\!.60$ for the wide image. 
The numbers of galaxies are 3717 and 6265, respectively.

To obtain stellar-mass functions of MODS samples, K09 performed SED fitting of the 
multiband photometry ($UBVizJHK$, 3.6 $\mu$m, 4.5 $\mu$m, and 5.8 $\mu$m) with population 
synthesis models.
We adopt the results with GALAXEV templates (Bruzual \& Charlot 2003) and the Salpeter initial mass
function (see K09 for more details).
In the present analysis, the near-infrared data (3.6$\mu$m) of Spitzer/IRAC 
are available for most of the sample galaxies (96\%), so that SED fitting is reasonably 
reliable for the photometry at rest-$z$ ($\lambda_\mathrm{eff} = 0.9 \mu$m) or 
shorter wavelengths out to redshift $\sim3$.

\section{Analyses}
\subsection{Stellar-mass Kormendy Relation}
The surface brightness $\mu_K$ (mag arcsec$^{-2}$) in $K$ band of the sample 
galaxies is defined as the brightness in two times the Kron radius, i.e.,
\begin{eqnarray}
\mu_K = m_K + 2.5log(\pi R^2),
\end{eqnarray}
where $m_K$ is total magnitude and $R$ is two times the Kron radius $r$ ($R=2r$) in units of arcsec.
For $m_K$, we use MAG\_AUTO obtained by SExtractor (Bertin \& Arnouts 1996), which gives the most 
precise estimate of the total magnitude of galaxies.
Kron radius is defined as the first moment,
\begin{eqnarray}
r =\frac{\Sigma r' I(r')}{\Sigma I(r')},
\end{eqnarray}
where $r'$ is isophotal radius and $I$ is flux at $r'$.
About 90\% or more of the total light is within $R$, irrespective
of galaxy shapes or redshifts (Kron 1980). 

The half-light radius and central SB of galaxies have been investigated in many previous 
studies.
However, in contrast, it is less reliable at higher redshifts in the present sample 
because of seeing effects.
Therefore, we study the total SB in $R$ instead.
Figure 1 shows $\mu_K$ versus stellar mass ($M_*$) of all galaxies broken into redshift bins.

\begin{figure}
\epsscale{1.0}
\plotone{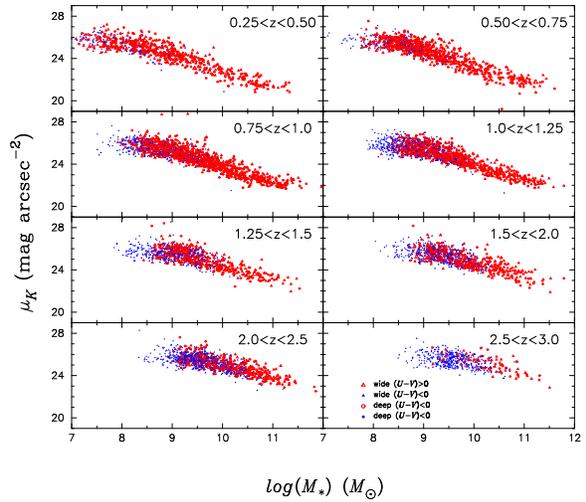}
\caption{Correlation of surface brightness in $K$ band ($\mu_K$) with stellar mass ($M_*$) 
of galaxies for each redshift bin. The galaxies are classified according to the color
 (rest-frame $U-V$) and the regions (wide and deep) as shown in the bottom right frame.
 }

\label{fig:fig1}
\end{figure}

\subsection{Point-like Sources}

In order to examine the error of SB, we first study the reliability of the Kron radius of 
galaxies.
Figure 2 shows the Kron radius derived by SExtractor versus $m_K$ for all MODS samples.
We draw a boundary line above the stars, which are spectroscopically confirmed.
Since spectroscopic data were not available for stars fainter than $\sim24$ mag 
in the present region, we examined the Kron radius using artificial stars.
The point sources were convolved with a gaussian point spread function (FWHM=$0\arcsec\!.5$), 
and then buried in the image with the same noise ($\sigma=30.9$ mag pixel$^{-2}$) with
that of the deep image.
SExtractor was applied to the artifacts in the same manner as for the MODS catalog.
The results are plotted over Figure 2.
The objects below the boundary in Figure 2 are considered as point-like or unresolved 
sources in the following.

\begin{figure}
\epsscale{0.8}
\plotone{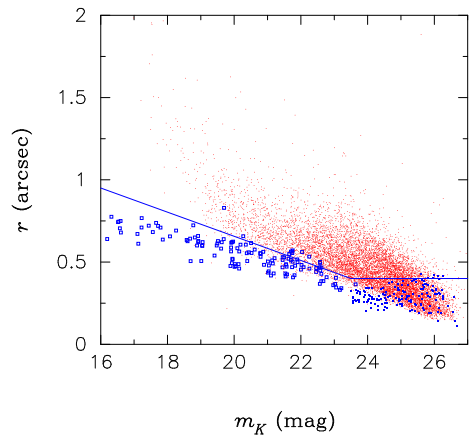}
\caption{Kron radius $r$ and total magnitude $m_K$ for the MODS catalog.
Squares show spectroscopically confirmed stars.
The results of SExtractor for artificial stars are plotted by filled circles.
The solid line delineates the upper boundary of point-like or unresolved sources.
 }

\label{fig:fig1}
\end{figure}

\subsection{Monte Carlo Study with Mock Galaxies}

Seeing and background noise strongly affect the observed Kron radius, especially 
for faint or small galaxies.
To examine the effect, we generated mock galaxies with a 1/4-law or exponential profile
and buried them in the simulated noise image.
The galaxies were randomly generated with various magnitudes and effective or scale lengths,
then analyzed with SExtractor in the same manner as for the MODS catalog.
The result is shown in Figure 3. 
The observed Kron radii $r$ are seriously affected by seeing for galaxies with 
an original Kron radius $r_0\lesssim0\arcsec\!.4$.
The observed SB of the mock galaxies with $r_0\lesssim0\arcsec\!.4$ is systematically
fainter than the originals.
We note that faint SB galaxies are strongly affected by background noise.
For example, the large scatter at $log(r/r_0)<0$ in Figure 3 is mainly due to low
SB galaxies with SB$\gtrsim26$ mag arcsec$^{-2}$.
Figure 3 also shows that the SB of galaxies with observed SB$>26$ mag arcsec$^{-2}$ is
unreliable.
Therefore, we will exclude the galaxies with $r\lesssim0\arcsec\!.4$ and/or SB$>26$ mag 
arcsec$^{-2}$ from the following discussion.

\begin{figure}
\epsscale{1.0}
\plotone{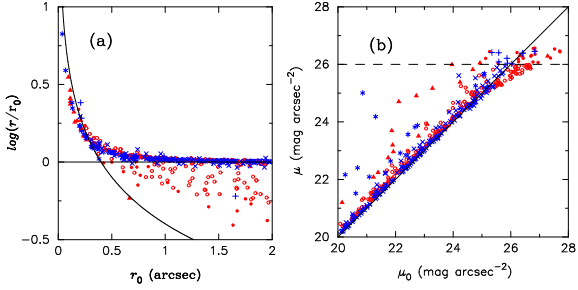}
\caption{(a) Kron radius and (b) surface brightness for mock galaxies.
The abscissa is the original values and the ordinate is those derived with SExtractor.
Crosses (blue) and circles (red) show galaxies with 1/4-law and exponential profiles, respectively.
The galaxies with $\mu>26$ are depicted with filled circles and pluses.
The solid curve in (a) is the observed Kron radius $r=0.4$.
The galaxies below the curve are noted with filled triangles and asterisks in (a) and (b).
The galaxies below the solid curve at  $r=0.4$ in (a) and above the dashed line at
$\mu=26$ are excluded from the present study.
}
\end{figure}

\begin{figure}
\epsscale{1.0}
\plotone{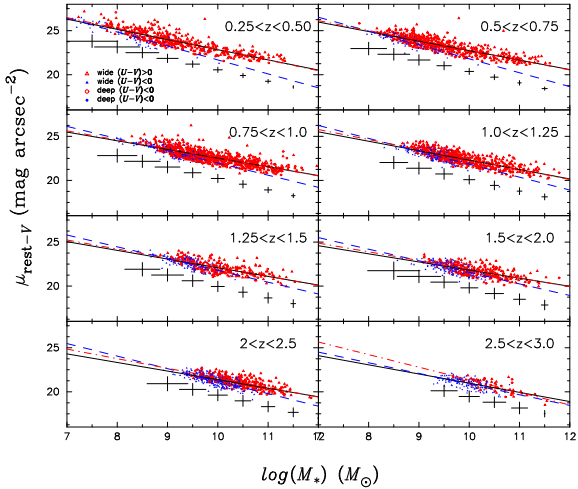}
\caption{Same as Figure 1, but for the rest-$V$ frame obtained with Equation 3.
Typical errors are depicted by crosses for mass bins.
Solid, dash, and dash-dot lines are the linear fits 
for all, blue ($U-V<0$), and red ($U-V\geq0$) samples, respectively.
}
\end{figure}

\begin{figure}
\epsscale{1.0}
\plotone{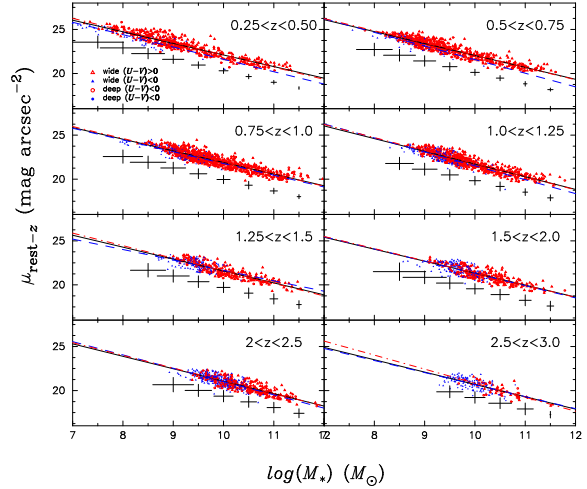}
\caption{Same as Figure 4, but for the rest-$z$ frame.
}
\end{figure}
\begin{figure}
\epsscale{1.0}
\plotone{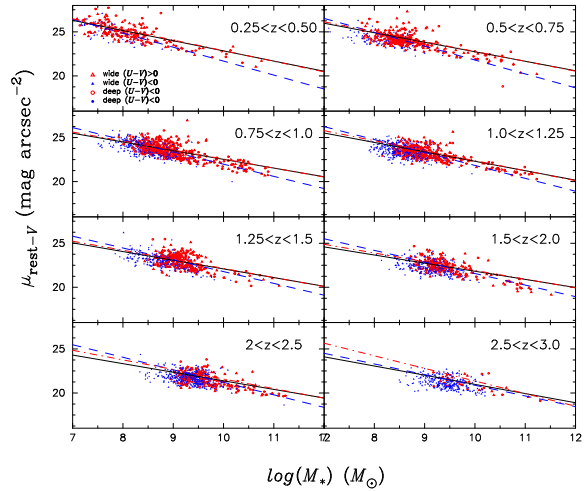}
\caption{Same as Figure 4, but for discarded galaxies.
The same regression lines as in Figure 4 are shown for reference.
}
\end{figure}

\subsection{$K-Corrections$}
Although a universal correlation independent on redshift is suggested in Figure 1,
when we discuss the evolution of the correlation as a function
of redshift, $k$-correction should be applied to the observed $m_K$ 
to compare them at the same rest-frame wavelengths.
The SB,  $\mu_\mathrm{rest}(\lambda)$ in a rest frame at wavelength $\lambda$,
is obtained from the observed data and the model SED, using the equation,
\begin{eqnarray}
\mu_\mathrm{rest}(\lambda) = m_K + k(\lambda)+2.5log(\pi R^2) - 2.5log(1+z)^3, 
\end{eqnarray}
where $k(\lambda)$ is the $k$-correction defined as the color $k(\lambda)=$ (rest-$V-K$) or (rest-$z-K$) 
of the best SED model fitted to the observations by the $\chi^2$ analysis (K09).

We apply the $k$-correction to $m_K$ for each sample galaxy to obtain the SB in the rest-$V$ and $z$ bands.
The last term $(1+z)^3$ is the correction for the dimming effect due to the cosmological expansion, 
after $m_K$ in AB system is reinstated by a factor of $(1+z)$.
Figures 4 and 5 shows the correlation of $M_*$ and SB in the rest-$V$ and $z$ frames, 
respectively.
The point-like galaxies ($r<0\arcsec\!.4$) and the galaxies with $\mu_K>26$ mag arcsec$^{-2}$ 
 are discarded from the analysis.
The discarded objects are depicted separately in Figure 6 for comparison.
Since the sizes of galaxies are suggestive of independence on wavelengths
in the local universe (e.g., Shen et al.\ 2003; Bernardi et al.\ 2003) and 
at high redshift (e.g., Cassata  et al\ 2009 at $z\sim2$), 
no $k$-correction for galaxy size is applied in the present analysis.

\subsection{$Error Estimate$}
The errors in stellar mass and $k$-correction mainly originated from 
SED fitting and the photometric errors of the observations.
In the course of $\chi^2$ fitting to obtain a best SED model with various parameters 
(e.g., star-formation time scale, photometric redshift for galaxies with no spectroscopic redshift available, 
age, extinction, and metallicity), the probability distributions of stellar mass and 
$k$-correction can be calculated, where the photometry and photometric-redshift errors 
are included.
The error for the observed SB is examined with mock galaxies in Figure 3 as a function 
of Kron radius. 
The photo-z error for galaxies with no spectroscopic redshift was studied in K09.
From these errors, we obtained  1-$\sigma$ errors for stellar mass and surface brightness 
for each galaxy in the rest-$V$ and rest-$z$ bands. 
The average errors for the stellar-mass bins are depicted in Figures 4 and 5.

\section{Results} \label{Results}

\begin{deluxetable}{ccrccccccc}
\tabletypesize{\scriptsize}
\tablecaption{Linear fit of the surface brightness in rest-$V$ and rest-$z$ frames to stellar mass of galaxies.\label{tbl-1}}
\tablewidth{0pt}
\tablehead{
\colhead{} & \colhead{} & \colhead{} & \colhead{} & \colhead{rest-$V$}  & \colhead{} & \colhead{} & \colhead{rest-$z$}
\\
\colhead{Redshift} &  \colhead{Sample} & \colhead{N} & \colhead{$a$}
& \colhead{$\mu_\mathrm{rest}^{10}$} & \colhead{$\sigma$} & \colhead{$a$} & \colhead{$\mu_\mathrm{rest}^{10}$} & \colhead{$\sigma$} 
\\
\colhead{} & \colhead{} & \colhead{} & \colhead{} & \colhead{(mag arcsec$^{-2}$)} & \colhead{} & \colhead{} & \colhead{(mag arcsec$^{-2}$)}
}
\startdata

$ 0.25\leq z<0.50 $  &    all                &   491  & $  -1.16 \pm0.03$  & $ 22.84 \pm 0.04 $ &  0.58 & $  -1.32 \pm0.02$  & $ 22.10 \pm 0.03 $ &  0.48\\
                     &   $U-V<0$             &    68  & $  -1.58 \pm0.09$  & $ 21.68 \pm 0.16 $ &  0.46 & $  -1.40 \pm0.08$  & $ 21.60 \pm 0.16 $ &  0.45\\
                     &   $U-V\geq0$          &   423  & $  -1.21 \pm0.03$  & $ 22.86 \pm 0.04 $ &  0.55 & $  -1.38 \pm0.03$  & $ 22.11 \pm 0.03 $ &  0.45\\
                     & $M_*>10^{10}M_\sun$\tablenotemark{a}   &    92  &                    & $ 22.23 \pm 0.04 $ &  0.41 &                    & $ 21.28 \pm 0.05 $ &  0.44\\
$ 0.50\leq z<0.75 $  &    all                &   626  & $  -1.08 \pm0.03$  & $ 22.74 \pm 0.03 $ &  0.52 & $  -1.35 \pm0.02$  & $ 22.05 \pm 0.02 $ &  0.43\\ 
                     &   $U-V<0$             &    83  & $  -1.58 \pm0.09$  & $ 21.80 \pm 0.12 $ &  0.39 & $  -1.55 \pm0.09$  & $ 21.57 \pm 0.12 $ &  0.38\\
                     &   $U-V\geq0$          &   543  & $  -1.12 \pm0.03$  & $ 22.77 \pm 0.03 $ &  0.51 & $  -1.39 \pm0.02$  & $ 22.07 \pm 0.02 $ &  0.42\\
                     & $M_*>10^{10}M_\sun$\tablenotemark{a}    &   144  &                    & $ 22.12 \pm 0.05 $ &  0.61 &                    & $ 21.22 \pm 0.05 $ &  0.58\\
$ 0.75\leq z<1.0  $  &    all                &   930  & $  -0.99 \pm0.02$  & $ 22.53 \pm 0.02 $ &  0.51 & $  -1.32 \pm0.02$  & $ 21.90 \pm 0.01 $ &  0.42\\
                     &   $U-V<0$             &   152  & $  -1.38 \pm0.08$  & $ 21.98 \pm 0.09 $ &  0.43 & $  -1.33 \pm0.08$  & $ 21.78 \pm 0.08 $ &  0.41\\
                     &   $U-V\geq0$          &   778  & $  -1.02 \pm0.03$  & $ 22.56 \pm 0.02 $ &  0.52 & $  -1.36 \pm0.02$  & $ 21.91 \pm 0.02 $ &  0.42\\
                     & $M_*>10^{10}M_\sun$\tablenotemark{a}    &   302  &                    & $ 21.96 \pm 0.03 $ &  0.61                      & $ 21.08 \pm 0.04 $ &  0.62\\
$ 1.0\leq z<1.25  $  &    all                &   698  & $  -1.07 \pm0.03$  & $ 22.30 \pm 0.02 $ &  0.49 & $  -1.44 \pm0.03$  & $ 21.72 \pm 0.02 $ &  0.44\\
                     &   $U-V<0$             &   184  & $  -1.47 \pm0.10$  & $ 21.84 \pm 0.07 $ &  0.45 & $  -1.58 \pm0.10$  & $ 21.52 \pm 0.08 $ &  0.47\\
                     &   $U-V\geq0$          &   514  & $  -1.13 \pm0.03$  & $ 22.37 \pm 0.02 $ &  0.48 & $  -1.48 \pm0.03$  & $ 21.76 \pm 0.02 $ &  0.42\\
                     & $M_*>10^{10}M_\sun$\tablenotemark{a}    &   201  &                    & $ 21.67 \pm 0.04 $ &  0.63 &                    & $ 20.83 \pm 0.04 $ &  0.63\\
$ 1.25\leq z<1.5  $  &    all                &   396  & $  -0.99 \pm0.04$  & $ 22.08 \pm 0.03 $ &  0.51 & $  -1.35 \pm0.04$  & $ 21.57 \pm 0.02 $ &  0.44\\
                     &   $U-V<0$             &   151  & $  -1.34 \pm0.11$  & $ 21.78 \pm 0.07 $ &  0.42 & $  -1.18 \pm0.11$  & $ 21.65 \pm 0.07 $ &  0.44\\
                     &   $U-V\geq0$          &   245  & $  -1.03 \pm0.06$  & $ 22.14 \pm 0.03 $ &  0.54 & $  -1.43 \pm0.05$  & $ 21.60 \pm 0.03 $ &  0.44\\
                     & $M_*>10^{10}M_\sun$\tablenotemark{a}    &   124  &                    & $ 21.65 \pm 0.06 $ &  0.62 &                    & $ 20.90 \pm 0.06 $ &  0.64\\
$ 1.5\leq z<2.0 $    &    all                &   430  & $  -0.92 \pm0.05$  & $ 21.82 \pm 0.02 $ &  0.51 & $  -1.36 \pm0.04$  & $ 21.33 \pm 0.02 $ &  0.47\\
                     &   $U-V<0$             &   174  & $  -1.31 \pm0.11$  & $ 21.56 \pm 0.05 $ &  0.45 & $  -1.39 \pm0.12$  & $ 21.30 \pm 0.06 $ &  0.51\\
                     &   $U-V\geq0$          &   256  & $  -0.99 \pm0.06$  & $ 21.91 \pm 0.04 $ &  0.52 & $  -1.38 \pm0.06$  & $ 21.35 \pm 0.03 $ &  0.45\\
                     & $M_*>10^{10}M_\sun$\tablenotemark{a}    &   177  &                    & $ 21.40 \pm 0.05 $ &  0.62 &                    & $ 20.66 \pm 0.05 $ &  0.64\\
$ 2.0\leq z<2.5 $    &    all                &   565  & $  -0.97 \pm0.04$  & $ 21.38 \pm 0.02 $ &  0.47 & $  -1.41 \pm0.04$  & $ 21.07 \pm 0.02 $ &  0.44\\
                     &   $U-V<0$             &   294  & $  -1.42 \pm0.07$  & $ 21.21 \pm 0.03 $ &  0.42 & $  -1.51 \pm0.08$  & $ 21.03 \pm 0.03 $ &  0.45\\
                     &   $U-V\geq0$          &   271  & $  -1.09 \pm0.06$  & $ 21.58 \pm 0.04 $ &  0.46 & $  -1.44 \pm0.05$  & $ 21.11 \pm 0.04 $ &  0.44\\
                     & $M_*>10^{10}M_\sun$\tablenotemark{a}    &   347  &                    & $ 20.91 \pm 0.03 $ &  0.61 &                    & $ 20.39 \pm 0.04 $ &  0.71\\
$ 2.5\leq z<3.0 $    &    all                &   184  & $  -1.04 \pm0.08$  & $ 20.98 \pm 0.04 $ &  0.48 & $  -1.40 \pm0.09$  & $ 20.70 \pm 0.04 $ &  0.51\\
                     &   $U-V<0$             &   127  & $  -1.21 \pm0.12$  & $ 20.88 \pm 0.04 $ &  0.44 & $  -1.37 \pm0.15$  & $ 20.68 \pm 0.05 $ &  0.53\\
                     &   $U-V\geq0$          &    57  & $  -1.43 \pm0.15$  & $ 21.36 \pm 0.10 $ &  0.48 & $  -1.59 \pm0.15$  & $ 20.84 \pm 0.09 $ &  0.46\\
                     & $M_*>10^{10}M_\sun$\tablenotemark{a}    &   104  &                    & $ 20.57 \pm 0.06 $ &  0.64 &                    & $ 20.14 \pm 0.07 $ &  0.72\\
                                                                                                                                                     
\enddata
\tablenotetext{a}{Average for galaxies with $M_*>10^{10}M_\sun$}
\end{deluxetable}

We obtain the least square fit of rest-$V$ (rest-$z$) SB 
to $M_*$ for the galaxies in Figure 4 (Figure 5) with a linear regression, 
\begin{eqnarray}
\mu_\mathrm{rest} = a (log M_* -10) + \mu_\mathrm{rest}^{10},
\end{eqnarray}
where $\mu_\mathrm{rest}^{10}$ is SB at $M_*=10^{10}M_\sun$ in the rest frame.

Since our image quality is not high enough for classifying the galaxies 
into morphological classes at high-z, we divided the samples into two groups according 
to rest-frame $U-V$ color and obtained again the linear fit for each group. 
Rest-frame $U-V=0$ corresponds to the color of A0 stars, so that the bluer group 
is supposed to be dominated by young and star-forming galaxies.
The results of the regression analysis with mean error are summarized in Table 1. 
The dispersion, $\sigma$, of the linear fit is listed in the last column.
(For galaxies with $M_*>10^{10}M_\sun$, the average SBs are obtained instead of 
linear fitting.)
Figure 7 shows $a$, the slope of Equation 4, as a function of redshift for
each group.
The figure indicates no noticeable systematic redshift dependence, 
except a hint  that the high redshift slope is slightly 
steeper than the values at lower redshift in rest-$V$ for the blue group, though the error is large.

The offset  $\mu_\mathrm{rest}^{10}$ is plotted in Figure 8 as a function of redshift for each sample.
Since the SBs depend on redshift for a given mass,
we fit $\mu_\mathrm{rest}^{10}$ in Figure 8 with redshift dependence,
\begin{eqnarray}
\mu_\mathrm{rest}^{10} = \mu_0 + 2.5 log(1+z)^n.
\end{eqnarray}
The results are shown in Figure 8 and Table 2.

The results, corrected for dust extinction, are also listed in Table 2, where the extinction 
is given from the best fit SED model by K09.
It should be noted that the dust extinction in galaxies is a less reliable parameter in SED 
fitting (K09).

\begin{deluxetable}{ccc}
\tabletypesize{\scriptsize}
\tablecaption{Redshift dependence of the surface brightness.\label{tbl-3}}
\tablewidth{0pt}
\tablehead{
\colhead{Sample} & \colhead{$n$} & \colhead{$\mu_{0}$} \\
\colhead{} & \colhead{} & \colhead{(mag arcsec$^{-2}$)} 
}
\startdata
\multicolumn{3}{c}{rest-$V$}               \\
\cline{1-3}
all                  & $ -1.74 \pm 0.13 $ & $23.63 \pm 0.12$ \\
$U-V\geq0$           & $ -1.46 \pm 0.07 $ & $23.49 \pm 0.07$  \\
$U-V<0$              & $ -0.78 \pm 0.25 $ & $22.28 \pm 0.24$  \\
$M_*>10^{10}M_\sun$  & $ -1.53 \pm 0.14 $ & $22.92 \pm 0.13$  \\
\cline{1-3}
\multicolumn{3}{c}{rest-$V$ (extinction corrected)}        \\
\cline{1-3}
all                  & $ -1.97 \pm 0.10 $ & $23.13 \pm 0.09$ \\
$U-V\geq0$           & $ -1.78 \pm 0.10 $ & $23.04 \pm 0.09$  \\
$U-V<0$              & $ -1.48 \pm 0.26 $ & $22.47 \pm 0.25$  \\
$M_*>10^{10}M_\sun$  & $ -1.85 \pm 0.10 $ & $22.24 \pm 0.10$  \\
\cline{1-3}
\multicolumn{3}{c}{rest-$z$}               \\
\cline{1-3}
all                  & $ -1.30 \pm 0.11 $ & $22.71 \pm 0.10$ \\
$U-V\geq0$           & $ -1.22 \pm 0.09 $ & $22.68 \pm 0.08$  \\
$U-V<0$              & $ -0.83 \pm 0.22 $ & $22.13 \pm 0.21$  \\
$M_*>10^{10}M_\sun$  & $ -1.05 \pm 0.09 $ & $21.75 \pm 0.09$  \\
\cline{1-3}
\multicolumn{3}{c}{rest-$z$ (extinction corrected)}               \\
\cline{1-3}
all                  & $ -1.42 \pm 0.10 $ & $22.44 \pm 0.09$ \\
$U-V\geq0$           & $ -1.38 \pm 0.09 $ & $22.43 \pm 0.09$  \\
$U-V<0$              & $ -1.21 \pm 0.24 $ & $22.24 \pm 0.23$  \\
$M_*>10^{10}M_\sun$  & $ -1.22 \pm 0.09 $ & $21.39 \pm 0.09$  \\
\enddata
\end{deluxetable}

\begin{figure}
\epsscale{1.0}
\plotone{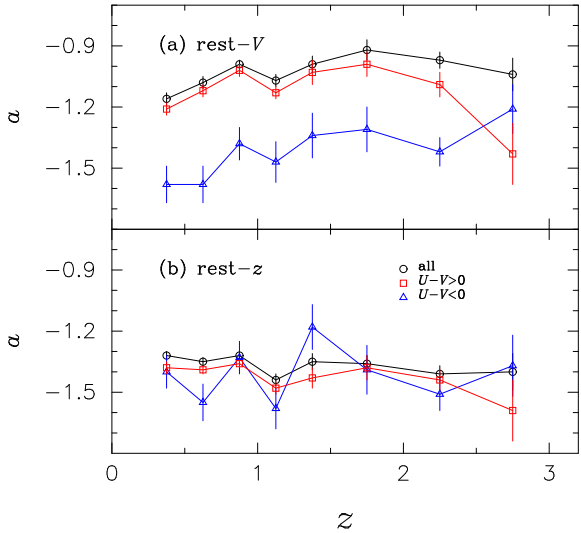}
\caption{Slope $a$ of the linear regression in Equation 4 as a function of 
redshift for each group in (a) rest-$V$ and (b) rest-$z$ frames.
}
\label{fig:fig7}
\end{figure}
                             
\begin{figure}
\epsscale{1.0}
\plotone{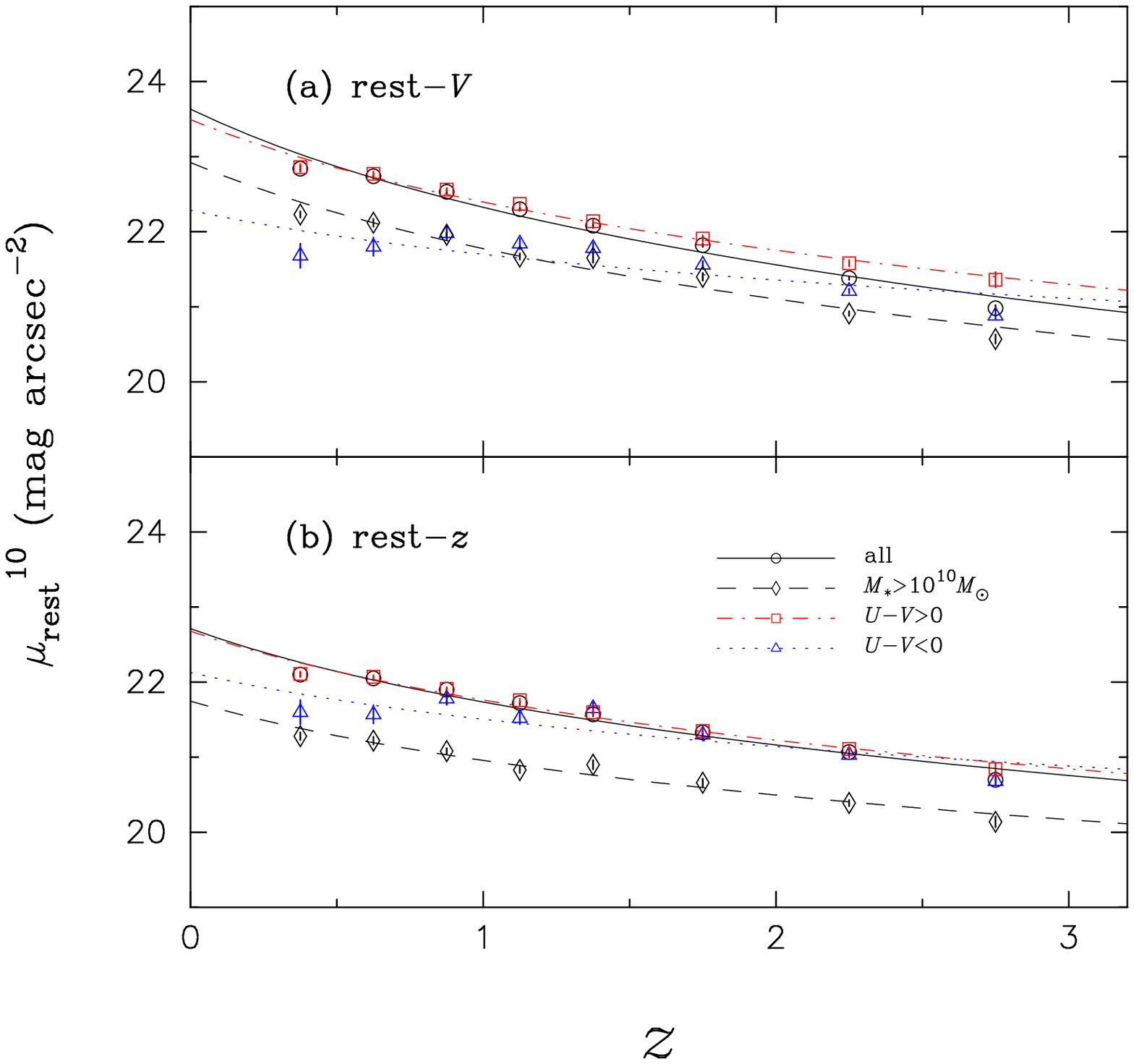}
\caption{Surface brightness ($\mu_\mathrm{rest}^{10}$) at $M_*=10^{10}M_\sun$ as a function of redshift.
Diamonds are the average SB for $M_*>10^{10}M_\sun$. 
Extinction is not corrected.
The line is the fitting result for each group (Table 2). 
}

\label{fig:fig8}
\end{figure}

\section{Discussion}
We have obtained the linear correlation between the SB and stellar mass of galaxies
in the GOODS-N region at $0.3<z<3$, which is an analogy of the Kormendy relation. 
The correlation has a nearly constant slope independent of redshift and the color of galaxies,
especially in the rest-$z$ frame.
Although the low-mass limit of the observations depends on redshift, the similar slopes of the 
linear regression in different redshift bins are suggestive of a universal relation between the SB 
and stellar mass of galaxies in a wide mass range.
The stellar system of galaxies keeps the tight correlation over the cosmic time.

Considering that the light in the rest-$z$ band is dominated by old stars and therefore traces the 
stellar mass, we note that the result should demonstrate a universal relation between surface stellar-mass 
density and stellar mass of galaxies.
This finding indicates that stellar mass in galaxies was build up over cosmic time 
in a similar manner, irrelevant to galaxy mass, as opposed to the scenario that 
massive galaxies mainly accumulated their stellar mass by major merging.
Dissipationless merging increases the size of galaxies, which leads 
to lower SB (Navarro 1990; Boylan-Kolchin, Ma, \& Quataert 2006; McIntosh et al.\ 2005; 
Damjanov et al.\ 2009; Nipoti et al.\ 2009).
If dry merging is a more important process for more massive galaxies, as conjectured 
from $\Lambda$CDM model, then the slope should become shallower at lower redshifts for 
massive galaxies.

Figure 8 shows the redshift dependence of $\mu_\mathrm{rest}^{10}$.
As a whole, SB is brighter at higher redshifts for a given mass.
The brightening obeys the relation with $n=-2.0\sim-0.8$ in rest-$V$ and $n=-1.4\sim-0.8$ 
in rest-$z$, depending on the subgroups and extinction correction used.
In Figure 9, we compare the observed SB of galaxies with the luminosity 
evolution, for a given stellar mass with no size change, for two extreme cases, i.e. a constant star 
formation (SF) model and an instantaneous-burst galaxy model formed at $z_\mathrm{f}=9$ (Bruzual \& 
Charlot 2003).
It is noted that we discuss the properties of galaxies with same stellar mass at any redshifts,
not tracing the evolution of mass or luminosity for a certain galaxy.
The SB evolution of red ($U-V>0$) and blue ($U-V<0$) samples is well reproduced by the 
luminosity evolution models of burst and constant SF models, respectively,
though SB is a little bluer than the models at higher redshift.
However, the models depend on various parameters (e.g., $z_\mathrm{f}$, metallicity, star-formation rate,
extinction, etc).
For example, lower metallicity at higher redshift makes the color bluer, which gives
an evolution more consistent with the observation. 
In any case, our result supports the scenario that galaxies accumulate stellar mass mainly by star formation.

\begin{figure}
\epsscale{1.0}
\plotone{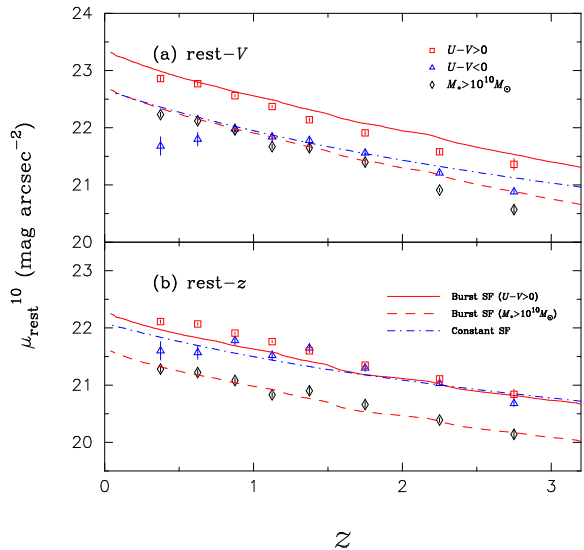}
\caption{Comparison of the observed surface brightness for blue, red, and massive
samples with the luminosity evolution, for a given stellar mass, with a constant star-formation model
(dash-dot line) and an instantaneous-burst galaxy model (solid and dash lines) formed 
at $z_\mathrm{r}=9$ (Bruzual \& Charlot 2003). 
The models are arbitrarily offset at the rest-$V$ band to fit the observations. 
}
\label{fig:fig9}
\end{figure}

The different selection criteria and analysis would not facilitate the direct comparison
of the present result with those of previous studies, which discussed luminosity evolution 
mainly in half-light radius based on images with a high spatial resolution.
The studies at high redshifts could be biased toward compact and massive galaxies.
In contrast, we focus on more general galaxies and studied the evolution of their 
SB based on the Kron radius for a given mass, making use of our deep imaging data,
which are favorable for diffuse galaxies.
Despite of the difference, the present results are consistent with previous studies.

For example, our result for the blue population supports the finding of Barden et al. (2005)
that the average surface brightness of disk galaxies increases with redshift
by about 1 mag to $z\sim1$ for galaxies.
Ferguson et al. (2004) and Akiyama et al. (2008) showed the size evolution of LBG galaxies 
at high redshifts, which indicates much stronger SB brightening than our results.
Although the blue group includes most LBGs in our catalog (Ichikawa et al. 2007),
our different selection criteria sample various populations of more general 
star-forming galaxies with low stellar mass, which would have more moderate evolution
(Trujillo et al. 2004; Trujillo et al. 2006).
The results for red massive groups are also in good agreement with those in
previous studies. 
Early-type galaxies were brighter by $1\sim2$ mag at $z\sim1$ (McIntosh et al. 2005)
and  $z\sim1.5$ (Damjanov et al. 2009),
which would correspond to our red group.

Trujillo et al.\ (2007) for galaxies with $M_*>10^{11}M_\sun$ and Toft et al.\ (2009) 
for those with $M_*>5\times 10^{10} M_\sun$ showed that massive galaxies were 
much smaller 
in the past than their local massive counterparts and the evolution was particularly strong 
for the highly concentrated objects, which are supposed to be high SB galaxies. 
Since our sample includes few such massive galaxies, we calculated the average
SB and its evolution in galaxies with $M_*>10^{10}M_\sun$ for
comparison.
The result is in good agreement with their results for resolved galaxies.
Their finding that the evolution is stronger for spheroid-like or quiescent galaxies
 would be supported by our result for the galaxies with the dust extinction corrected (Table 2).

Finally we emphasize that the present study has not only confirmed the previous studies
on the SB evolution of galaxies, but also demonstrated that the brightening dates back 
to $z\sim3$.
We should also mention about the point-like or unresolved galaxies ($r<0\arcsec\!.4$),
which are excluded from the analysis for the present stellar-mass Kormendy relation.
For example, early-type galaxies (Damjanov et al.\ 2009) 
and compact galaxies (van Dokkum et al.\ 2008) tended to be a factor 2--5 smaller 
in half light radius than local counterparts of similar mass.
The SBs of galaxies at the faint-end limit of each redshift bin in Figure 6 are 
seriously affected by background noise, while more massive galaxies would not.
The real SB of the massive galaxies, which are well above the SB limit but unresolved 
in the present observation, could be much brighter.
The samples of e.g., Akiyama et al. (2008), Trujillo et al. (2007), and Muzzin et al.\ (2009) would 
be such galaxies as those we did not resolve with our poorer image size.
If they are observed with higher spatial resolution, they could be plotted further below 
the regression lines in Figures 4 and 5.
Therefore, the stellar-mass Kormendy relation presented in this paper would rather show a 
lower boundary of the correlation of stellar mass and SB of galaxies; lower SB galaxies are 
prohibited over the boundary.
 
\acknowledgments

This work has been supported in part by a Grant-in-Aid for Scientific Research (21244012) 
of the Ministry of Education, Culture, Sports, Science and Technology in Japan.
We thank Ramsey Lundock for careful reading of the manuscript.

\end{document}